\documentclass[letterpaper, onecolumn, 10pt]{article}
\usepackage[left=2.5cm,right=2.5cm,top=3cm,bottom=3cm]{geometry}

\usepackage[utf8x]{inputenc}
\usepackage[spanish]{babel}
\usepackage{url}
\usepackage{microtype}
\usepackage{graphicx}
\usepackage{booktabs}
\usepackage{tabulary}

\usepackage{setspace}
\addto\captions{%

}


\usepackage[unicode=true]{hyperref}

\hypersetup{breaklinks=true,
            bookmarks=true,
            pdfauthor={Eduardo Graells-Garrido},
            pdftitle={Ornitología Virtual},
            colorlinks=true,
            urlcolor=blue,
            linkcolor=magenta,
            pdfborder={0 0 0}}

\title{Ornitología Virtual: Caracterizando a \#Chile en Twitter}
\author{Eduardo Graells-Garrido
\thanks{E. Graells-Garrido es parte del Web Research Group en Universitat Pompeu Fabra, Barcelona, España. Correo: \url{eduard.graells@upf.edu.}}
}

\date{}

\begin{document}
\maketitle

\begin{abstract}
Este artículo presenta un análisis de los \emph{tweets} recolectados el 28 de Octubre de 2012, en el contexto de las elecciones municipales de 2012 en Chile. 
Dicho análisis se realiza mediante una metodología basada en literatura previa, en particular en técnicas de recuperación de la información y de análisis de espacios de información. 
Como resultado, se determinan: 
\emph{1)} características demográficas básicas de la población virtual chilena, incluyendo su distribución geográfica,
\emph{2)} el contenido que caracteriza a cada región, y cómo fluye información entre regiones,
y \emph{3)} el grado de representatividad de la población virtual participante en el evento con respecto a la población física.
Se determina que la muestra obtenida es representativa de la población en términos de distribución geográfica, que el centralismo que afecta al país se ve reflejado en Twitter, y que, a pesar de los sesgos poblacionales, es posible identificar el contenido que caracteriza a cada región.
Se finaliza con una discusión de las implicaciones y conclusiones prácticas de este trabajo, así como futuras aplicaciones.

\emph{Palabras clave}---Redes Sociales, Caracterización de la Población, Interacción Humano-Computador, Recuperación de la Información, Geolocalización.
\end{abstract}

\section{Introducción}

Chile es un país con una gran penetración de tecnologías, tanto a nivel de \emph{hardware} como de \emph{software}. 
Ejemplos abundan: es un país donde la cantidad de telefonos móviles supera a la población \cite{subtelChile}, 
una gran cantidad de servicios estatales cuenta con procesos informáticos modernos y expeditos\footnote{Como ejemplo, considerar el portal \emph{ChileAtiende} en \url{http://www.chileatiende.cl/}.}, 
y su población se ha caracterizado por una gran presencia en diferentes redes sociales \cite{socnetlatam}. 
Este último punto motiva este trabajo, puesto que la actividad en redes sociales ha dejado de ser exclusiva de usuarios y usuarias avanzadas o \emph{early adopters}: como base, se asume que lo que sucede en redes sociales, en especial en aquellas con un perfil de contenido público como Twitter\footnote{\url{https://twitter.com}}, ya es parte de la conversación de ``gente común y corriente'', incluso de aquellos y aquellas que no participan activamente en la red social pero que sí están pendientes de lo que sucede en ella. 

Por otro lado, a diferencia del mundo físico, donde existen limitaciones geográficas que dividen a la población, en el mundo virtual dichas barreras no están presentes. 
Por tanto, se facilita el contacto entre personas de lugares alejados, en particular en un país como Chile. 
Así, este artículo se enfoca en la red social Twitter y cómo parte de la población chilena participó en un evento de importancia nacional: las elecciones municipales efectuadas el 28 de Octubre de 2012 \cite{wiki:chile2012elections}. Este evento es de interés para análisis puesto que fue relevante en todo el territorio, a su vez que el contenido generado es de alto interés local, por lo que estudiar su contenido permite comprender cómo es usado Twitter en cada región de Chile. 

Twitter es una red social implícitamente pública, que utiliza el símil de aves cantando al aire libre con el de personas \emph{twitteando} sus opiniones sobre lo que está sucendiendo en el momento. 
Inspirándose en la ornitología, el estudio de las aves, en este artículo se presenta un primer estudio que permite contextualizar la participación de chilenos y chilenas en Twitter, a modo de ornitología virtual, considerando los siguientes ejes de análisis:
\emph{1)} Características demográficas de los usuarios y usuarias participantes, en particular: sexo y ubicación geográfica, así como variables adicionales como tiempo de ingreso en la red social y biografías de las personas,
\emph{2)} El contenido del evento generado por la población virtual a través de los tweets publicados,
\emph{3)} La caracterización de las regiones del país en base a métricas de actividad y de contenido en los tweets.

\subsection{Preguntas de Investigación}

El objetivo principal de este artículo es realizar una primera caracterización contextual de la población chilena en Twitter.
Así, las preguntas de investigación a responder en este artículo son:

\begin{enumerate}
\item ¿Cómo se puede caracterizar la población virtual chilena en Twitter?
\item ¿Cómo determinar qué es lo que caracteriza el contenido de cada región del país? 
\end{enumerate}

\subsection{Contribuciones}

Este artículo presenta dos contribuciones. 
La primera es una caracterización de la relación entre un subconjunto de la población virtual chilena presente en Twitter, en la cual se encuentra que la población estudiada es espacialmente representativa de la población física chilena, que existe un fuerte sesgo hacia la población masculina y hacia la población con alto nivel educacional, y que el centralismo del país se ve reflejado. 
La segunda es una metodología basada en técnicas de recuperación de la información \cite{baeza2011modern} para mostrar que es posible, a pesar de los sesgos poblacionales, encontrar el contenido que caracteriza a cada región independientemente de su tamaño en términos de población.

\section{Marco Teórico}

Existen dos áreas de investigación que han inspirado profundamente este proyecto. 
La primera es la caracterización de Twitter y sus usuarios y usuarias. 
La segunda es la caracterización del espacio web de un país.

\subsection{Twitter}
Twitter es una red social que tiene propiedades de medios de comunicación \cite{kwak2010twitter}, es decir, la conectividad entre personas es central pero a su vez la red cumple con roles de difusión de información.
En este trabajo Twitter es analizada desde la caracterización demográfica de las personas que participan en la red, área de investigación que se ha llevado a cabo en gran parte con la población de los Estados Unidos de América, donde existe un fuerte sesgo de población masculina, las ciudades con mayor población están sobre-representadas y las ciudades con menor población están infra-representadas \cite{mislove2011understanding}. 

En Twitter las personas no ingresan datos estructurados sobre sí mismas, puesto que gran parte de la información es de texto libre (i.e., la ubicación se describe mediante palabras y una misma ubicación puede tener diferentes nombres). 
A pesar de esta libertad de ingreso de la información es factible analizar la ubicación geográfica de las personas que participan en la red social.
Un trabajo realizado en Estados Unidos determinó que el $66\%$ de las cuentas ingresa una ubicación geográfica válida \cite{hecht2011tweets}.

El poder clasificar la población de la red social en base a la geografía permite estudiar diferencias culturales \cite{poblete2011all}, aunque éstas se han enfocado en las diferencias entre países, dejando de lado las diferencias locales, potencialmente interesantes en un país con  diversidad cultural y geográfica como Chile.
También existen análisis inspirados en la sociología \cite{quercia2012social}, en los que se ha encontrado que las redes de cada persona (\emph{ego-networks}) tienen una fuerte componente local en el aspecto geográfico, que la emotividad de las personas tiene correlación con las estructuras de sus redes, y que los agentes sociales son aquellos que \emph{twittean} sobre temas diversos, o bien, aquellos que no son mono-temáticos.
 
En este trabajo se asume que existen diferencias detectables entre regiones de Chile, de modo de poder caracterizar el país a nivel local, considerándose cada región del país como unidad básica de análisis.

\subsection{Espacios Web Nacionales}
La segunda área es la caracterización del espacio web de un país. 
Esta caracterización consiste en determinar la estructura de la conexión mediante enlaces entre los sitios web que están en la red de cada país \cite{baeza2007characterization}, y cómo evoluciona dicha estructura con el tiempo \cite{graells2008evolution}. 
Sin embargo, los enlaces no son la única manera en la cual las personas ingresan a un sitio web -- la búsqueda es otra, particularmente importante hoy, puesto que prácticamente cualquier persona sabe utilizar un buscador. En este aspecto, se ha caracterizado cómo interactúan diferentes países \cite{baeza2009geographical}, teniendo en cuenta desde donde se busca (país de origen de la consulta) hasta donde se llega (país que alberga el contenido que se consume). 
Esta misma perspectiva se puede aplicar a Twitter, al ver como los países intercambian información entre sí \cite{kulshrestha2012geographic}. 

En este trabajo se toma el concepto de caracterización a nivel nacional, de modo que se expone cómo diferentes variables se distribuyen geográficamente, y cómo diferentes regiones están enlazadas entre sí mediante interacciones, definiendo así el flujo de información entre ellas.

\section{Metodología}

\subsection{Recolección de Datos}

Los tweets son recolectados utilizando la \emph{Streaming API} de Twitter \footnote{Application Programming Interface -- \url{https://dev.twitter.com/docs/streaming-apis}}. Esta API recibe un conjunto de términos, como \emph{hashtags}, palabras comunes o menciones a otras cuentas, y entrega en tiempo real los tweets asociados a los términos de búsqueda utilizados. 

\subsection{Características de la Población}

La caracterización de la población está basada en las ideas de Mislove et al. \cite{mislove2011understanding} y Hecht et al. \cite{hecht2011tweets}, es decir, se utiliza la información reportada por los usuarios y usuarias en sus perfiles para determinar su sexo y ubicación geográfica. También se considera la fecha en que las personas se han estado registrado en la red social, y se analiza el vocabulario empleado por las personas en sus biografías. 

\subsubsection{Sexo}

Para determinar el sexo de una persona se utiliza el nombre reportado por ésta, utilizando listas de nombres conocidos. 
En particular, se inició el estudio utilizando las listas disponibles en Wikipedia\footnote{\url{https://es.wikipedia.org/wiki/Categor\%C3\%ADa:Nombres_masculinos} y \url{https://es.wikipedia.org/wiki/Categor\%C3\%ADa:Nombres_femeninos}}. Después se inspeccionó manualmente la base de datos en búsqueda de nombres comunes que no hayan sido clasificados, con el fin de agregarlos a las correspondientes listas y realizar la clasificación nuevamente. En total se cuenta con $1522$ nombres masculinos y $1330$ nombres femeninos.

\subsubsection{Distribución Geográfica}

En Twitter, cada persona puede ingresar un texto en el cual menciona su ubicación. Este nombre se geolocaliza, es decir, se trata de asignar a una ubicación conocida. 
Para esto, se ha generado una lista de topónimos basada en una lista oficial del Ministerio del Interior \cite{minint2010}. 
Esta base de datos contiene las relaciones \emph{comuna} $\rightarrow$ \emph{provincia} $\rightarrow$ \emph{región} $\rightarrow$ \emph{país}. Para generar la lista se han utilizado las siguientes plantillas:

\begin{enumerate}\itemsep0pt
  \item \emph{comuna}.
  \item \emph{provincia}.
  \item \emph{comuna}, \emph{provincia}.
  \item \emph{provincia}, \emph{región}.
  \item \emph{región}.
  \item \emph{comuna}, \emph{país}.
  \item \emph{provincia}, \emph{país}.
  \item \emph{región}, \emph{país}.
  \item \emph{comuna} de \emph{país}.
  \item \emph{provincia} de \emph{país}.
  \item \emph{región} de \emph{país}.
  \item \emph{país}.
\end{enumerate}

Una vez generada la lista de topónimos, se inspeccionó la base de datos en busca de nombres comunes repetidos que no hayan sido generados por el método propuesto, y se realizó una clasificación manual de éstos. En total se tienen $1978$ topónimos para diferentes lugares de Chile, utilizados para geolocalizar usuarios y usuarias.

\subsubsection{Biografías}

En los perfiles existe un campo de texto llamado ``biografía'', en el cual las personas ingresan un texto breve para describirse a sí mismas si lo desean. 
Para cada persona clasificada como hombre o mujer se ha separado el texto de su biografía en múltiples palabras simplificadas (sin tildes y en minúsculas), y se han removido palabras funcionales (\emph{stopwords}). 
Así, para cada persona se tiene un conjunto de palabras. 
Cada uno de estos conjuntos se utiliza para crear un grafo de co-ocurrencias, en el cual cada nodo representa a una palabra y dos nodos están conectados si las dos palabras correspondientes aparecen en el mismo conjunto o biografía. 
El peso de cada arista es la fracción de biografías en las que aparecen ambas palabras. 
Para tener solamente palabras representativas, el grafo es filtrado de acuerdo al peso de las aristas: se dejan solamente las aristas dentro del $0.1\%$ superior en términos de peso. 
Luego, en el grafo filtrado se calcula \emph{Pagerank} \cite{page1999pagerank}, un algoritmo iterativo que estima la importancia de un nodo en términos de las conexiones que tiene. Al aplicar dicho algoritmo se establece un orden de importancia de las palabras más representativas.

Para cada palabra se puede calcular la tendencia con respecto al sexo de los usuarios. La tendencia se define de la siguiente manera:
$$tendency(w) = {{male\_users(w) - female_\_users(w)} \over total\_users(w)}$$
donde $male\_users(w)$ es la cantidad de hombres que incluye la palabra $w$ en sus biografíaas, mientras que $female\_users(w)$ es la cantidad de mujeres que incluye la palabra en sus biografías. 
Una palabra con tendencia $1$ es usada solamente por hombres, mientras que una palabra con tendencia $-1$ es usada solamente por mujeres. Una tendencia $0$ indica una palabra neutral.

\subsubsection{Tiempo}

Se considera la fecha en que las personas se han unido a la red social Twitter, de modo de construir una serie temporal que muestre el volumen diario de registros. En esta serie temporal se ha aplicado un algoritmo de reconocimiento de \emph{peaks} \cite{du2006improved}, de los cuales se considera solamente el decil superior en términos de volumen de registro.

Habiendo identificado los \emph{peaks} más importantes, se procede a buscar una explicación enciclopédica que permita descubrir el contexto histórico en el cual se han dado esos registros. La búsqueda se ha realizado en Wikipedia, teniendo como argumento que si un evento importante ocurrió alrededor de esos \emph{peaks}, dicho evento debiese estar documentado en la enciclopedia, y a su vez, si la fecha no aparece en la enciclopedia, se puede argumentar que en esos días no sucedió nada relevante como para ser incluido en ella. Se utiliza Wikipedia por ser de acceso libre y abierto. Existen antecedentes que indican que es tan precisa como enciclopedias tradicionales \cite{rosenzweig2006can}.

\subsection{Contenido del Evento}

A través del contenido generado por usuarios y usuarias participantes en el evento se busca determinar si hay diferencias en contenido y en comportamiento entre regiones. En particular, se considera la cantidad de tweets publicados por región como una medida de comportamiento, y el texto publicado desde cada región como una característica de contenido. También son considerados los tweets con coordenadas geográficas y las interacciones a nivel regional.

\subsubsection{Volumen de Tweets}

Para analizar la cantidad o volumen de tweets, por cada región se genera una serie temporal que contiene la cantidad de tweets publicada cada $5$ minutos. Luego, las series temporales son normalizadas, ya que las diferencias de población provocan que el volumen de tweets difiera entre una región y otra. Para normalizar las series temporales se calcula el máximo \emph{peak} de cada serie, y cada elemento de cada serie es dividido por el \emph{peak} correspondiente. Así, cada serie temporal contiene elementos que varían entre $0$ (sin tweets publicados) y $1$ (máximo de tweets publicados), por lo que es posible comparar qué tan diferentes o similares son entre sí en términos de volumen.

\subsubsection{Texto y \emph{Hashtags}}

Es posible que las \emph{hashtags} y menciones más populares a nivel nacional estén sesgadas por las regiones. Para corregir ese sesgo, se utiliza el modelo vectorial \cite{salton1975vector,baeza2011modern} para representar y organizar los tweets. En este modelo cada región del país es considerada como un documento cuyo contenido es el conjunto total de tweets emitidos desde ella. Cada documento es representado como un vector disperso:
$$R_{i} = [w_{0,i}, w_{1,i}, \ldots, w_{n,i}]$$
donde $R_{i}$ es una región, y $w_{j,i}$ es el peso de la palabra de vocabulario $j$ en la region $i$. 
En particular se utiliza un peso conocido como TF-IDF (\emph{term frequency-inverse document frequency}), que representa las veces que ha aparecido la palabra $j$ dentro del documento (\emph{term frequency}) ajustada por la cantidad de documentos en los que aparece dicha palabra (\emph{inverse document frequency}) de acuerdo a la siguiente formula:
$$TFIDF(w, r, R) = freq(w, r) \times log {|R| \over |r \in R: w \in r|}$$
donde $w$ es una palabra, $r$ es una región, $R$ es el conjunto de regiones y $freq$ es la cantidad de tweets originados en $r$ que contienen a $w$.
De este modo, las palabras que aparecen en todas o en varias regiones tienen menos peso dentro de un documento-región que una palabra que solamente aparece dentro de solo un documento-región. 
Esta fórmula permite filtrar palabras comunes entre regiones para así encontrar el vocabulario que caracteriza a cada una de ellas.

\subsubsection{Interacciones a Nivel Regional}

Las interacciones a nivel regional han sido modeladas a través de una matriz origen-destino definida de la siguiente manera:
$$M_{i,j} = mentions(R_{i}, R_{j})$$
donde $mentions$ indica la cantidad de tweets originados en la región $R_{i}$ (es decir, cuyo autor o autora pertenece a esa región) que mencionan a una o más cuentas en la región $R_{j}$. De esta manera se propone una medida del flujo de información entre regiones.

\subsubsection{Tweets con Información Geográfica}

Los tweets publicados desde \emph{smartphones}, \emph{tablets} y otros dispositivos con GPS, contienen coordenadas geográficas (latitud y longitud) que permiten conocer la ubicación desde la cual el tweet fue publicado\footnote{Se recomienda ver \url{https://blog.twitter.com/2013/the-geography-of-tweets}, un conjunto de mapas de ciudades confeccionados utilizando solamente este tipo de información.}. Se explora el volumen de tweets de acuerdo a sus coordenadas geográficas y a la representación de la densidad tecnológica de una ciudad en base a dicho volumen.

\section{Población Virtual}

El espacio de información que se analiza es un conjunto de $498594$ tweets y $270028$ retweets, publicados por $173077$ cuentas de usuario el 28 de Octubre de 2012 entre las $10$ A.M. y  medianoche, en el contexto de las elecciones municipales que se llevaron a cabo en Chile \cite{wiki:chile2012elections}. 
La Tabla \ref{table:information_space} muestra cifras básicas que caracterizan la colección de tweets obtenidos, luego de una etapa de filtro para excluir información que no es relevante al evento o al país\footnote{Tweets excluidos: 1) aquellos relacionados con eventos que sucedieron el mismo día pero que no estaban relacionados con \texttt{\#municipales2012}, 2) tweets en idiomas distintos al castellano, 3) tweets generados en el extranjero y no relacionados con Chile.}. Este evento se identificó con la hashtag \texttt{\#municipales2012}, aunque también se recolectaron tweets que contuviesen hashtags relacionadas\footnote{A lo largo del día se agregaron nuevas hashtags a monitorear de acuerdo a las noticias relevantes.}, nombres de candidatos, ciudades y cuentas relacionadas (por ejemplo, cuentas de candidatos conocidos). 
La Tabla \ref{table:query} muestra algunos términos de consulta utilizados para recolectar tweets.

\begin{table}[p]
\centering
\begin{tabular}{lc}
  \toprule
  Variable & Cantidad \\
  \midrule
  Tweets & $498594$ \\
  ReTweets & $270028$ \\
  Participantes & $173077$ \\
  Vocabulario & $289032$ \\
  \bottomrule
\end{tabular}
\caption{Características básicas de la muestra obtenida desde Twitter para el evento \texttt{\#municipales2012}. La cantidad de tweets no considera retweets. El vocabulario incluye todas las cadenas alfanuméricas encontradas, incluyendo \emph{hashtags} y menciones. Fuente: elaboración propia.}
\label{table:information_space}
\end{table}

\begin{table}[p]
\centering
\footnotesize
\begin{tabular}{l}
\toprule
Query  \\
\midrule
\texttt{H}: \texttt{\#municipales2012, \#túdecides, \#yovote}\\
\texttt{V}: \texttt{elecciones municipales, elección, abstención} \\
\texttt{V}: \texttt{voté, voten, votamos, \ldots} \\
\texttt{V}: \texttt{vocal de mesa, mesa, urna, deber cívico}\\
\texttt{V}: \texttt{alcade, alcadesa, alcaldía, concejal} \\
\texttt{P}: \texttt{concertación, alianza por chile, servel} \\
\texttt{P}: \texttt{labbe, errazuriz, zalaquett, tohá, sabat, \ldots} \\
\texttt{L}: \texttt{chile, santiago, concepción, valparaíso, \ldots} \\
\bottomrule
\end{tabular}
\caption{Términos de consulta empleados para monitorear el evento y descargar tweets. Leyenda: \emph{H} (\emph{hashtags}), \emph{V} (vocabulario), \emph{P} (política) y \emph{L} (ubicaciones). Fuente: elaboración propia.}
\label{table:query}
\end{table}

\subsection{Sexo}

La Tabla \ref{table:user_types} indica la cantidad de personas clasificadas por sexo basándose en el primer nombre reportado por usuarios y usuarias en sus perfiles. 
Un $68\%$ de las cuentas puede ser clasificada: $39.6\%$ pertecene a hombres y $28.4\%$ pertenece a mujeres. Las cuentas no determinadas ($32\%$) corresponden a los siguientes casos:
\emph{1)} personas que utilizan nombres que no estaban presentes en nuestras listas, 
\emph{2)} personas con nombres ambiguos, como Alexis,
\emph{3)} cuentas asociadas a medios de comunicación, empresas u otras entidades,
y \emph{4)} cuentas que utilizan nombres ficticios o falsos que impiden un reconocimiento automático.  
Se esperaría que hombres y mujeres tengan igual presencia en la red, pero, al igual que en resultados obtenidos en Estados Unidos \cite{mislove2011understanding}, en esta muestra de la población virtual chilena hay un marcado sesgo hacia la población masculina.

\begin{table}[p]
\centering
\begin{tabular}{lc}
\toprule
Sexo & \# Personas \\
\midrule
Hombre & $68583$ ($39.6\%$)\\
Mujer & $49123$ ($28.4\%$)\\
No Determinado & $55371$ ($32.0\%$)\\
\bottomrule
\end{tabular}
\caption{Cantidad de personas clasificas por sexo según su primer nombre, de acuerdo a lo reportado en sus perfiles de Twitter. Fuente: elaboración propia.}
\label{table:user_types}
\end{table}

\subsection{Distribución Geográfica}

\begin{table}[p]
\centering
\begin{tabular}{lcc}
\toprule
Nivel & \# Personas & \# Tweets\\
\midrule
Pa\'is & $17929$ ($10.4\%)$        & $98458$ ($12.8\%)$\\
Regi\'on & $341$ ($0.1\%)$     & $2637$  ($0.3\%)$\\
Provincia & $1842$ ($1.1\%)$      & $12541$ ($1.6\%)$\\
Ciudad/Comuna & $52961$ ($30.6\%)$ & $325326$ ($42.3\%)$\\
No Determinado & $100004$ ($57.8\%)$ & $329660$($42.9\%)$\\
\bottomrule
\end{tabular}
\caption{Distribución de la clasificación de personas y los tweets publicados según el nivel administrativo de la ubicación detectada. Los porcentajes han sido redondeados a un decimal. Fuente: elaboración propia.}
\label{table:location_levels}
\end{table}

La Tabla \ref{table:location_levels} muestra la cantidad de personas clasificadas de acuerdo al nivel geográfico detectado en la geolocalización basada en topónimos. 
El porcentaje de cuentas no clasificadas geográficamente es alto: $57.8\%$. A pesar de esto, el $42.2\%$ que sí es clasificado genera un $57.1\%$ de los tweets recolectados, por lo que se considera que el muestreo de cuentas geolocalizadas es suficiente para poder realizar el análisis de las secciones siguientes. 
Un $35\%$ de las cuentas participantes no reporta ubicación.

\begin{figure}[htbp]
\centering
\includegraphics[width=0.9\textwidth]{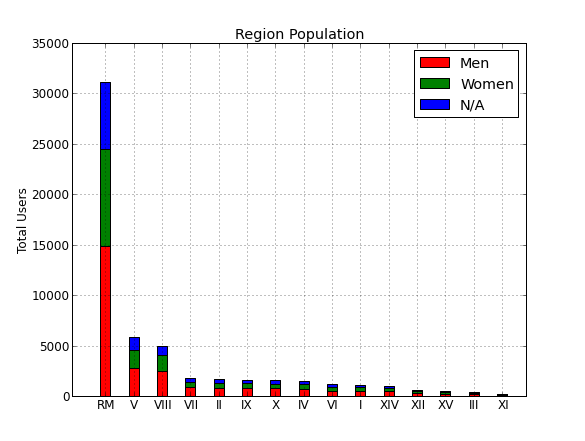}
\caption{Distribución de población de hombres y mujeres en las regiones de Chile. Fuente: elaboración propia.
\label{fig:region_population}}
\end{figure}

\begin{figure}[htbp]
\centering
\includegraphics[width=0.9\textwidth]{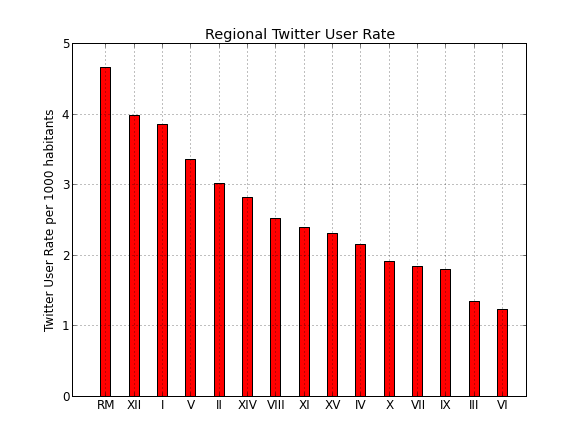}
\caption{Tasa de \emph{twitteros} y \emph{twitteras} por cada $1000$ habitantes en las regiones de Chile. Fuente: elaboración propia.
\label{fig:region_population_rate}}
\end{figure}

La Figura \ref{fig:region_population} muestra la cantidad de personas clasificadas para cada región del país, y la Figura \ref{fig:region_population_rate} muestra la tasa de \emph{twitteros} y \emph{twitteras} por cada $1,000$ habitantes en cada región. 
Considerando que la tasa de \emph{twitteros} y \emph{twitteras} no presenta diferencias en grados de magnitud, y que la \emph{correlación de pearson}\footnote{La correlación de pearson $r \in [-1, 1]$ es una medida de la relación lineal entre dos variables aleatorias. Un valor de $1$ indica relación lineal, un valor de $-1$ indica relación lineal inversa, y un valor de $0$ indica que no hay relación.} entre los logaritmos de las poblaciones física y virtual de las regiones es $0.95$ ($p < 0.001$), es posible indicar que la muestra de la población virtual es espacialmente representativa con respecto a la población física.

\subsection{Biografías}
\label{sec:bios}

Las $50$ palabras más importantes, junto a sus respectivas tendencias y frecuencias, se encuentran en la Tabla \ref{table:top_bio_keywords}. 
Se observa que la palabra más importante, y a su vez más frecuente, es \emph{estudiante}, utilizada por $15\%$ de las personas y con una ligera tendencia masculina. 
Se observan otras palabras relacionadas con estudios y profesiones, como \emph{universidad} (pos. 5), \emph{ingeniero} (pos. 9), \emph{ingenieria} (pos. 12), \emph{civil} (pos. 13), \emph{comercial} (pos. 16), \emph{periodista} (pos. 17), \emph{ing} (abreviacion de \emph{ingenieria}, pos. 18), \emph{derecho} (pos. 19), entre otras, aunque no se observan palabras que vinculen a los estudiantes con liceos o colegios. 
También hay palabras que describen o vinculan gustos: \emph{amante} (pos. 2), \emph{musica} (pos. 6), \emph{colo} (pos. 15), \emph{rock} (pos. 42), \emph{cine} (pos. 47). 
Algunas mujeres, que tienen menor presencia, utilizan \emph{hija} (pos. 24), \emph{mujer} (pos. 26), \emph{mama} (pos. 33) y \emph{madre} (pos. 34), mientras que los hombres no utilizan \emph{hombre} pero sí \emph{padre} (pos. 23) e \emph{hijo} (pos. 40).

Los resultados sugieren que la población en Twitter que participó en este evento tiene un alto nivel educacional y es, en general, mayor de edad.

\begin{table}[p]
\centering
\footnotesize
\begin{tabular}{llcc}
\toprule
{} & Palabra & Tendencia & Porcentaje de Personas \\
\midrule
1  &  estudiante  & $0.14$ & $15.03\%$ \\
2  &  amante  & $0.16$ & $7.31\%$ \\
3  &  vida  & $-0.09$ & $7.67\%$ \\
4  &  chile  & $0.30$ & $6.32\%$ \\
5  &  universidad  & $0.22$ & $4.54\%$ \\
6  &  musica  & $0.14$ & $4.54\%$ \\
7  &  corazon  & $0.26$ & $4.35\%$ \\
8  &  amo  & $-0.35$ & $3.94\%$ \\
9  &  ingeniero  & $0.74$ & $3.86\%$ \\
10  &  hincha  & $0.64$ & $2.99\%$ \\
11  &  feliz  & $-0.30$ & $4.06\%$ \\
12  &  ingenieria  & $0.48$ & $2.57\%$ \\
13  &  civil  & $0.60$ & $2.50\%$ \\
14  &  futbol  & $0.68$ & $2.43\%$ \\
15  &  colo  & $0.63$ & $1.89\%$ \\
16  &  comercial  & $0.43$ & $2.42\%$ \\
17  &  periodista  & $-0.02$ & $3.70\%$ \\
18  &  ing  & $0.50$ & $2.40\%$ \\
19  &  derecho  & $0.24$ & $2.54\%$ \\
20  &  com  & $0.34$ & $2.37\%$ \\
21  &  familia  & $0.00$ & $2.22\%$ \\
22  &  fanatico  & $0.96$ & $1.96\%$ \\
23  &  padre  & $0.91$ & $1.92\%$ \\
24  &  hija  & $-0.62$ & $1.31\%$ \\
25  &  social  & $0.01$ & $2.01\%$ \\
26  &  mujer  & $-0.52$ & $1.78\%$ \\
27  &  educacion  & $0.07$ & $1.42\%$ \\
28  &  trabajo  & $0.19$ & $2.26\%$ \\
29  &  industrial  & $0.55$ & $1.40\%$ \\
30  &  amigos  & $0.15$ & $1.50\%$ \\
31  &  encanta  & $-0.33$ & $1.58\%$ \\
32  &  hijos  & $0.11$ & $1.28\%$ \\
33  &  mama  & $-0.93$ & $2.10\%$ \\
34  &  madre  & $-0.89$ & $1.66\%$ \\
35  &  informatica  & $0.64$ & $1.10\%$ \\
36  &  vivo  & $0.12$ & $1.70\%$ \\
37  &  catolica  & $0.17$ & $0.96\%$ \\
38  &  mundo  & $0.06$ & $2.09\%$ \\
39  &  candidato  & $0.97$ & $0.39\%$ \\
40  &  hijo  & $0.61$ & $1.26\%$ \\
41  &  love  & $-0.48$ & $0.93\%$ \\
42  &  rock  & $0.48$ & $1.14\%$ \\
43  &  estudio  & $0.12$ & $1.59\%$ \\
44  &  amiga  & $-0.98$ & $0.78\%$ \\
45  &  historia  & $0.35$ & $1.05\%$ \\
46  &  dios  & $-0.11$ & $1.22\%$ \\
47  &  cine  & $0.21$ & $1.19\%$ \\
48  &  naturaleza  & $0.01$ & $1.31\%$ \\
49  &  profesor  & $0.95$ & $1.36\%$ \\
50  &  enamorada  & $-0.96$ & $1.20\%$ \\
\bottomrule
\end{tabular}
\caption{Las $50$ palabras más importantes en las biografías de las personas. Fuente: elaboración propia.}
\label{table:top_bio_keywords}
\end{table}

\subsection{Tiempo}

\begin{figure}[htbp]
\centering
\includegraphics[width=\textwidth]{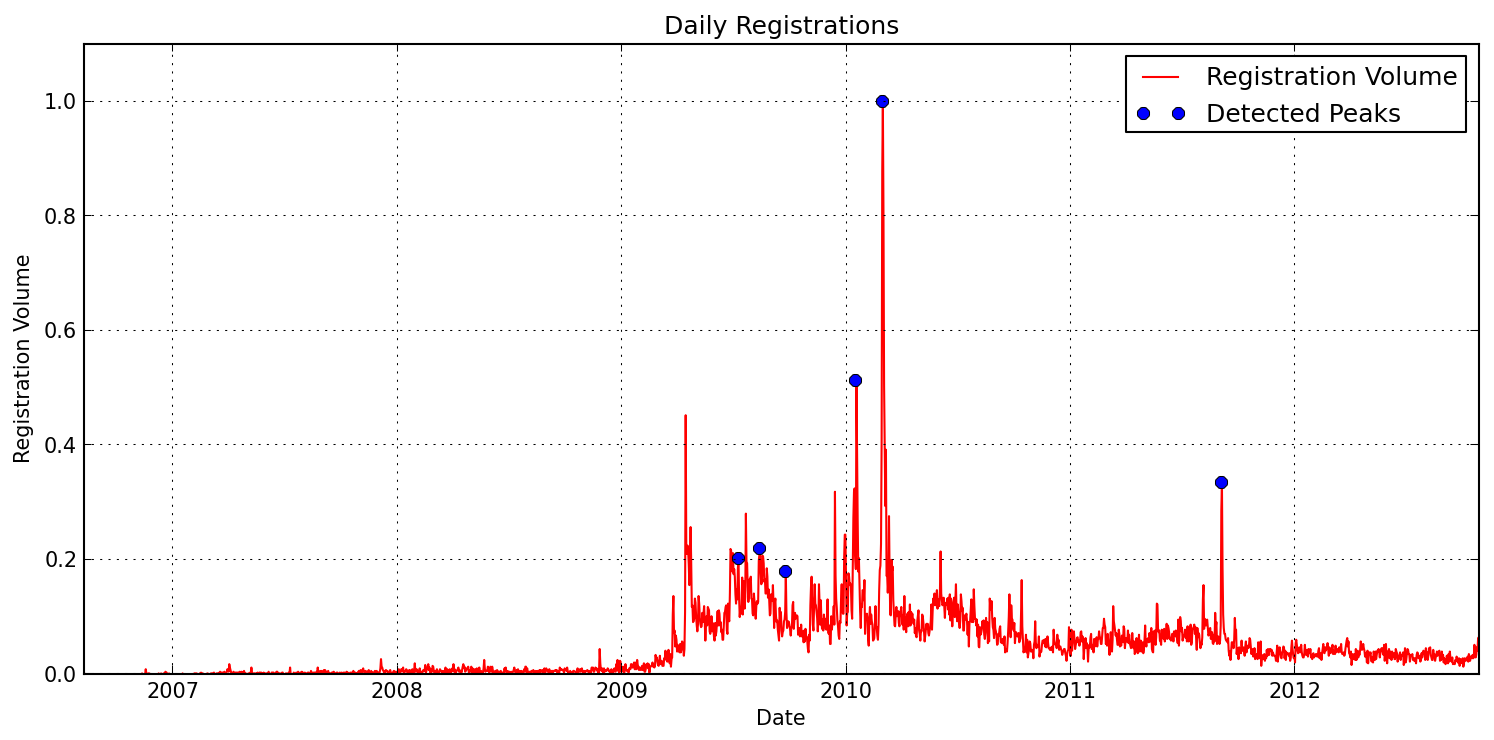}
\caption{Serie temporal que representa la cantidad de chilenos y chilenas registradas en Twitter cada día. Fuente: elaboración propia.}
\label{fig:registration_timeseries}
\end{figure}

La Figura \ref{fig:registration_timeseries} muestra la cantidad de registros diarios en Twitter. 
Se observa que desde el inicio de la red social\footnote{Twitter inició sus servicios el año 2006.} es posible encontrar personas chilenas, aunque su uso era minoritario: la población estaba totalmente compuesta por \emph{early adopters}. 
El auge en el uso comenzó el año 2009, en particular, el 29/05/2009 el $10\%$ de las cuentas recolectadas estaba registrada en la red social, y sólo tres meses después, el 13/08/2009, ya se había duplicado, es decir, el $20\%$ de las cuentas ya se había registrado. 
La Tabla \ref{table:registration_decils} muestra la fecha correspondiente a cada decil de población en términos de fecha de registro.

\begin{table}[p]
\centering
\footnotesize
\begin{tabular}{lcc}
\toprule
Fecha & Porcentaje & Días Transcurridos desde Decil Anterior \\
\midrule
09/08/2006 & $0$ & -- \\
20/05/2009 & $10$ & $1015$ \\
13/08/2009 & $20$ & $85$ \\
21/11/2009 & $30$ & $100$ \\
07/02/2010 & $40$ & $78$ \\
07/04/2010 & $50$ & $59$ \\
19/07/2010 & $60$ & $103$ \\
29/12/2010 & $70$ & $163$ \\
23/06/2011 & $80$ & $176$ \\
24/12/2011 & $90$ & $184$ \\
28/10/2012 & $100$ & $308$ \\
\bottomrule
\end{tabular}
\caption{Fechas de registro de cada decil de la población virtual chilena en términos de antigüedad. Fuente: elaboración propia.}
\label{table:registration_decils}
\end{table}

Luego de aplicar el algoritmo de detección de \emph{peaks} previamente especificado, se obtienen $175$ días clasificados como \emph{peaks}. 
El alto número de días se debe a que la serie temporal tiene mucho ruido y por ende el número de máximos locales es alto. 
Para filtrar dichos días se ha dejado el decil de registros más alto en términos de volumen de registros, obteniendo así $6$ $peaks$,  graficados como puntos sobre la serie temporal de la Figura \ref{fig:registration_timeseries}. 
Las fechas correspondientes y su posible explicación de acuerdo a Wikipedia se discuten a continuación:

\begin{itemize}
  \item \textbf{09/07/2009, 13/08/2009, 2009-09-24}: desde fines de Abril hasta Julio hubo una pandemia de gripe A (H1N1) \cite{wiki:pandemiaGripe}. El 12 de Agosto fue asesinado un comunero mapuche en La Araucanía \cite{wiki:conflictoMapuche}, y el 23 de Septiembre se realizó el primer debate televisado entre candidatos de las elecciones presidenciales de 2009 \cite{wiki:chile2009elections}. 
  \item \textbf{2010-01-17}: \emph{``Segunda vuelta de la elección presidencial. Se enfrentan los candidatos Sebastián Piñera (Coalición por el Cambio) y Eduardo Frei (Concertación de Partidos por la Democracia), resultando ganador el primero con el 51,6\% de los votos válidamente emitidos''} \cite{chile2010}.
  \item \textbf{01/03/2010}: \emph{``27 de febrero: Terremoto 8,8° en la escala de Richter sacude a la zona centro-sur del país, su epicentro fue Cobquecura Región del Biobio. El terremoto y el consecuente tsunami dejaron 525 muertos identificados y 25 personas desaparecidas.5 A raíz del desastre se cancela la noche de clausura del Festival de Viña del Mar''} \cite{chile2010}. El día 1 de Marzo de 2013 es el que tiene más registros en la red social. Si bien el terremoto sucedió dos días antes, el acceso a Internet había sido limitado. Estudios al respecto se pueden encontrar en \cite{niclabsterremoto,Sepulveda:2010:SCI:1978002.1978068}. La actividad generada en la red social inspiró investigaciones sobre credibilidad en la dispersión de rumores \cite{castillo2011information}.
  \item \textbf{04/09/2011}: \emph{``3 de septiembre: El presidente Sebastián Piñera recibió en el Palacio de La Moneda a dirigentes estudiantiles para lograr una solución a las demandas del movimiento estudiantil''} \cite{chile2011}. En la Sección \ref{sec:bios} se observa que muchas personas se definen como estudiantes, por lo que se puede especular que el movimiento estudiantil de 2011 \cite{wiki:movimientoEstudiantil} influenció el registro de nuevas personas en la red social.
\end{itemize}

\section{Contenido del Evento}

\subsection{Volumen de Tweets}

\begin{figure}[htbp]
\centering

\includegraphics[width=\textwidth]{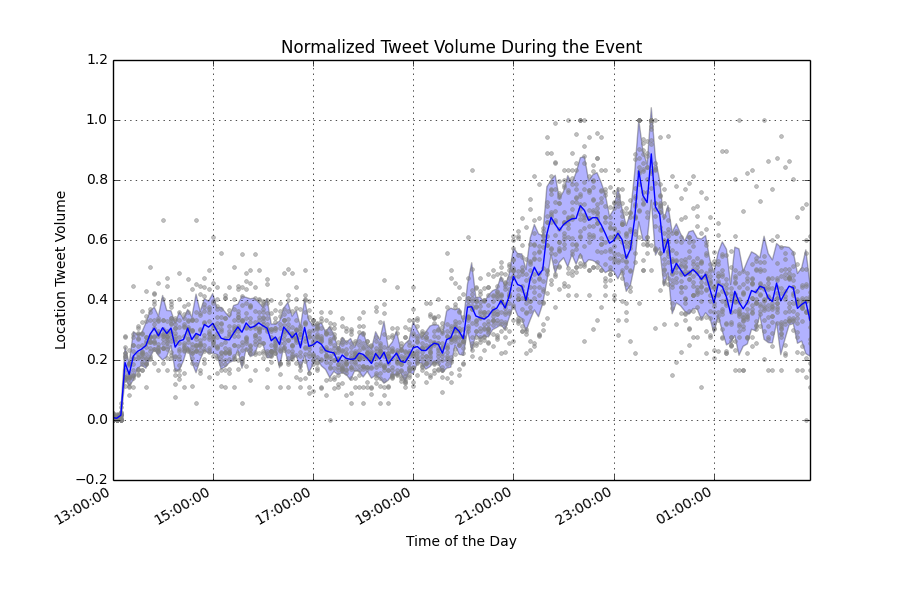}
\caption{Cantidad de tweets publicada cada $5$ minutos durante el evento para cada región de Chile. La hora desplegada es GMT. Fuente: elaboración propia.}
\label{fig:tweet_volume_regions}
\end{figure}

La Figura \ref{fig:tweet_volume_regions} muestra los tweets generados desde cada región del país, en un conteo por intervalos de $5$ minutos. Se pone énfasis en que son tweets generados puesto que ello no implica que el contenido del tweet se relacione con la región desde la cual fue emitido. 
En la Figura \ref{fig:tweet_volume_regions} se observa el comportamiento de las regiones en este aspecto: cada punto representa el volumen normalizado de tweets en una región particular, la curva representa el promedio de los volúmenes de tweets y la área alrededor de la curva representa la desviación estándar. 
Se observa que en general el comportamiento fue similar, es decir, tiene la misma estructura entre regiones: el volumen comienza a subir a medida que empiezan los recuentos de votos, y tiene un repunte cuando hay sorpresas en las votaciones, como la derrota de \emph{Cristian Labbé} (anterior alcalde de la comuna de Providencia en Santiago, identificado con la hashtag \texttt{\#labbe}) ante \emph{Josefa Errazuriz} (de cuenta \texttt{@josefaerrazuriz} en Twitter). 
Esta derrota fue controversial y tuvo gran repercusión en la discusión: además de \texttt{\#labbe} y \texttt{@josefaerrazuriz}, destacan las \emph{hashtags} \texttt{\#chaolabbe}, \texttt{\#trabajosparalabbe} y \texttt{\#providencia} entre las más populares del país, como se verá en la siguiente sección.

\subsection{Texto y \emph{Hashtags}}

\begin{table}[p]
\centering
\scriptsize
\begin{tabular}{lc|lc}
\toprule
Menci\'on & \#Tweets & Hashtag & \#Tweets \\
\midrule
\texttt{@biobio} & $14283$ & \texttt{\#municipales2012} & $115710$ \\
\texttt{@josefaerrazuriz} & $9541$ & \texttt{\#tudecides} & $19230$ \\
\texttt{@cooperativa} & $9400$ & \texttt{\#labbe} & $9565$ \\
\texttt{@tv\_mauricio} & $6813$ & \texttt{\#chile} & $5306$ \\
\texttt{@cnnchile} & $5947$ & \texttt{\#nunoa} & $5250$ \\
\texttt{@mayafernandeza} & $5484$ & \texttt{\#valdiviacl} & $4701$ \\
\texttt{@patricionavia} & $5474$ & \texttt{\#iquique} & $4603$ \\
\texttt{@envivo\_fm} & $4605$ & \texttt{\#providencia} & $4560$ \\
\texttt{@copano} & $4574$ & \texttt{\#cooperativa} & $3451$ \\
\texttt{@24horastvn} & $4203$ & \texttt{\#yovote} & $3015$ \\
\texttt{@chilevision} & $4179$ & \texttt{\#yovoto} & $2885$ \\
\texttt{@christianpino} & $4087$ & \texttt{\#zalaquett} & $2581$ \\
\texttt{@dondatos} & $4008$ & \texttt{\#antofagasta} & $2369$ \\
\texttt{@carolina\_toha} & $3212$ & \texttt{\#santiago} & $2231$ \\
\texttt{@carolaurrejola} & $2985$ & \texttt{\#rmv} & $2204$ \\
\texttt{@guatonsalinas} & $2947$ & \texttt{\#muncipales2012} & $2127$ \\
\texttt{@tele13online} & $2825$ & \texttt{\#desobedientes} & $2089$ \\
\texttt{@tvn} & $2735$ & \texttt{\#magallanes} & $1978$ \\
\texttt{@difamadores} & $2594$ & \texttt{\#chaolabbe} & $1893$ \\
\texttt{@fernandopaulsen} & $2589$ & \texttt{\#elecciones2012} & $1837$ \\
\texttt{@giorgiojackson} & $2466$ & \texttt{\#proyeccionescnnbiobio} & $1701$ \\
\texttt{@emol} & $2440$ & \texttt{\#recoleta} & $1608$ \\
\texttt{@marcoporchile} & $2414$ & \texttt{\#trabajosparalabbe} & $1503$ \\
\texttt{@adnradiochile} & $2411$ & \texttt{\#municipales} & $1495$ \\
\texttt{@mwaissbluth} & $2362$ & \texttt{\#temuco} & $1451$ \\
\bottomrule
\end{tabular}
\caption{Menciones y \emph{hashtags} más populares en el evento para toda la población. Fuente: elaboración propia.}
\label{table:mentions_and_hashtags}
\end{table}

Al analizar el contenido textual generado de los tweets, se observa que se han mencionado $77940$ nombres de usuario diferentes (incluyendo cuentas no existentes o mal escritas) y que se han utilizado $21,891$ \emph{hashtags} distintas. 
A su vez, un $37.51\%$ de los tweets contiene \emph{hashtags}. 
La Tabla \ref{table:mentions_and_hashtags} contiene las $25$ menciones y \emph{hashtags} más populares a nivel nacional. 
Se observa prevalencia de cuentas relacionadas con figuras políticas (como \texttt{@josefaerrazuriz} en pos. 2 y \texttt{@mayafernandeza} en pos. 6), periodistas (como \texttt{@tv\_mauricio} en pos. 4 y \texttt{@copano} en pos. 9) y medios de comunicación (como \texttt{@biobio} en pos. 1 y \texttt{@cooperativa} en pos. 3), y que respecto a las \emph{hashtags}, hay relacionadas con el evento (como \texttt{\#municipales2012} en pos. 1 y \texttt{\#tudecides} en pos. 2), su coyuntura (como \texttt{\#labbe} en pos. 3 y \texttt{\#zalaquett} en pos. 12), y con lugares (como \texttt{\#nunoa} en pos. 5 y \texttt{\#valdiviacl} en pos. 6).

La Tabla \ref{table:region_relevant_hashtags} muestra las \emph{hashtags}, menciones y palabras más relevantes para cada región al aplicar el modelo vectorial con TF-IDF, es decir, los elementos con más peso de los vectores dispersos de cada región. 
Esto asegura que dichas palabras sean las más discriminativas a la hora de decidir a que región pertenecen. 
Dentro de las \emph{hashtags} se observa una mayor presencia de nombres de lugares (como \texttt{\#copiapo} en III Región, \texttt{\#laserena} en IV Región, \texttt{\#concon} en V Región, \texttt{\#melipilla} en RM y \texttt{\#coyhaique} en XI).  
Dentro de las menciones se observan cuentas pertenecientes a personalidades locales (como \texttt{@gervoyparedesr}, alcalde de una provincia en la X Región, \texttt{@emilioboccazzi}, alcalde de una provincia en la XII Región), medios locales (como \texttt{@radiorancagua} en VI Región, \texttt{@antofagastatv} en II Región y \texttt{@biobiotv} en VIII Región) y cuentas locales (como \texttt{@felipelmagno} en XV Región). 
Dentro de las palabras ``normales'' se encuentran mayoritariamente lugares y apellidos. 
De este modo, se muestra que el modelo vectorial permite encontrar, por ejemplo, tweets relacionados con medios locales sin necesidad de conocer las cuentas de dichos medios. 
Por contraste, no aparecen medios populares, como \texttt{@biobio} y \texttt{@cooperativa}, los más populares de acuerdo a la Tabla \ref{table:mentions_and_hashtags}. Esto se debe a que son cuentas mencionadas a nivel nacional y por ende no son menciones que identifiquen la ubicación del tweet que las contiene, mientras que un tweet que menciona a \texttt{@desiertofm} es altamente probable que provenga o que hable de la II Región, o bien un tweet que contiene el apellido \texttt{valcarce} probablemente hable de a la XV Región, refiriéndose a la entonces candidata \emph{Ximena Valcarce}.

\begin{table}[p]
\centering
\scriptsize
\begin{tabulary}{\textwidth}{L|L}
\toprule
Reg. & Hashtags, Palabras y Menciones Relevantes \\
\midrule
III \hspace{1cm} & \texttt{copiapo \#copiapo \#atacama cicardini \#freirina cid maglio lopez caldera freirina atacama \#tierraamarilla \#copiapoelige12 chanaral amarilla marcos @radiomaray vallenar @cecy\_landia barahona \#vallenar \#caldera @radionostalgica \#tierra volta} \\
\midrule IV & \texttt{\#laserena @elobservatodo \#ovalle ovalle jacob \#coquimbo \#combarbala combarbala \#municipaleseldia @eldia\_cl @laserena\_chile lobos hanna renteria pizarro galleguillos \#sismo navea \#eldia\_cl moira castagneto harufe yusta coquimbo jarufe} \\
\midrule VII & \texttt{talca \#talca \#curico \#talcavota \#linares curico parral maule @logikafm linares \#constitucion hermosilla renteria alexis \#parral vielma tilleria @juancastrotalca sepulveda basualto castro @ancoafm fital \#romeral @futurafmoficial} \\
\midrule V & \texttt{sumonte urenda concon valdovinos \#concon municipalesfm \#valparaiso \#municipalesfm @soysanantonio valpo quillota \#quillota \#valpo \#vina llay vinadelmar quilpue pedofilo @ucvradio olmue vinambres @cotyreginato \#vinadelmar stevenson coty} \\
\midrule RM & \texttt{renca \#melipilla huechuraba quilicura \#actualidad macul conchali \#chilevotaust \#macul \#puentealto vittori pudahuel melipilla @envivo\_fm actualidad \#maipu \#pudahuel pontificia pradanos codina \#quilicura @123noticias vitacura \#maya @duranbaronti} \\
\midrule XI & \texttt{huala coyhaique \#coyhaique \#aysen aysen acevedo luperciano \#decisionmunicipal @hualacoyhaique @radiosantamaria portales catalan @trapanandavivo @radiokuykuy @pcuevasm \#chilechico cochrane marcia raphael @patagonia970 atton @josceron @mauriciomnoz @intendencia11 2990} \\
\midrule X & \texttt{\#puertomontt \#osorno @gervoyparedesr \#puertovaras \#dalcahue bertin tejeda miramar \#municipalescastro2012 llanquihue \#ancud purranque gervoy \#chiloe melipulli osorno \#chonchi ancud berger @jaimebertin quellon \#quellon montt \#purranque brahm} \\
\midrule XII & \texttt{puq \#puqvota \#puq \#municipalesmag arenas boccazzi \#polartv \#puntaarenas @emilioboccazzi \#votapuq polar sahr magallanes \#magallanes eterovic \#pinguinotv @radiomagallanes arcos mimica punta karelovic amar aguilante \#mesa125 bocazzi} \\
\midrule I & \texttt{\#iquique @elboyaldia tarapaca dubost \#soria myrta \#altohospicio nortino tarapacatv @\_natasilva \#tarapaca cejas twitcam @iquique\_tweet @red\_mivoz @sociopedro @tarapacatv itv \#itv @hombrederadio @cristianjamett @jmarroquing @radiopaulina @hugo\_gutierrez\_ @dorysharder} \\
\midrule XV & \texttt{\#aricavota \#arica valcarce rocafull alamo abdala stancic chinga chang @nairayatiri @soyarica @felipelmagno urrutia @tremenatojones @iusarmijo \#urrutia chipana azapa retenido saucache lissette sierra ariquenos rocaful @muqsaki} \\
\midrule XIV & \texttt{\#valdiviacl \#lu2012 \#uach panguipulli \#valdivia uach valdivianos amtmann @valdivianos @omar\_sabat @carlosamtmann ilabaca mafil n\_n paillaco salesiano @radioportales @eduardoarcosa \#launionchile @coxismo bamtmann \#riobueno \#panguipulli kunstmann kohler} \\
\midrule IX & \texttt{\#araucania \#temuco \#araucaniaelige araucania @elperiodicocl huenchumilla carahue pucon \#pucon \#eleccionespucon tiburcio villarrica \#padrelascasas \#carahue shemagh saudi bilal @laopinon @chavezpau @proaraucania @uatv\_noticias edita pitrufquen @beckeralcalde group} \\
\midrule VIII & \texttt{zarzar \#chillan bernucci chillan \#coronel @gastonsaavedra penco @biobiotv \#concepcion chiguayante armstrong hualpen @cristian\_quiroz cristian\_quiroz @armstrongconce fabiola \#talcahuano conce \#chiguayante orbe experimental quiroz \#sanpedrodelapaz lota rivas} \\
\midrule II & \texttt{\#antofagasta hernando antofagastatv antofagasta @antofagastatv \#karenrojo \#calama @karenrojov karen portilla marcela tocopilla sabella calama @velasquezcalama mejillones velasquez @canuo rancho adaro \#tocopilla lopez @marcelahernando dostora @desiertofm} \\
\midrule VI & \texttt{rancagua \#rancagua arellano soto machali @alcaldesoto llancao lisboa graneros @radiorancagua \#machali pichilemu aurora rengo sismos jadell \#sanfernando segovia @tatianamerino @carlosampuero \#graneros mostazal rgua \#pichilemu berwart} \\
\bottomrule
\end{tabulary}
\caption{Palabras, menciones y \emph{hashtags} más características para cada región del país, detectadas utilizando el modelo vectorial con pesos TF-IDF. Fuente: elaboración propia.}
\label{table:region_relevant_hashtags}
\end{table}

\subsection{Interacciones}
\label{sec:interactions}

\begin{figure}[htbp]
\centering
\includegraphics[width=0.55\textwidth]{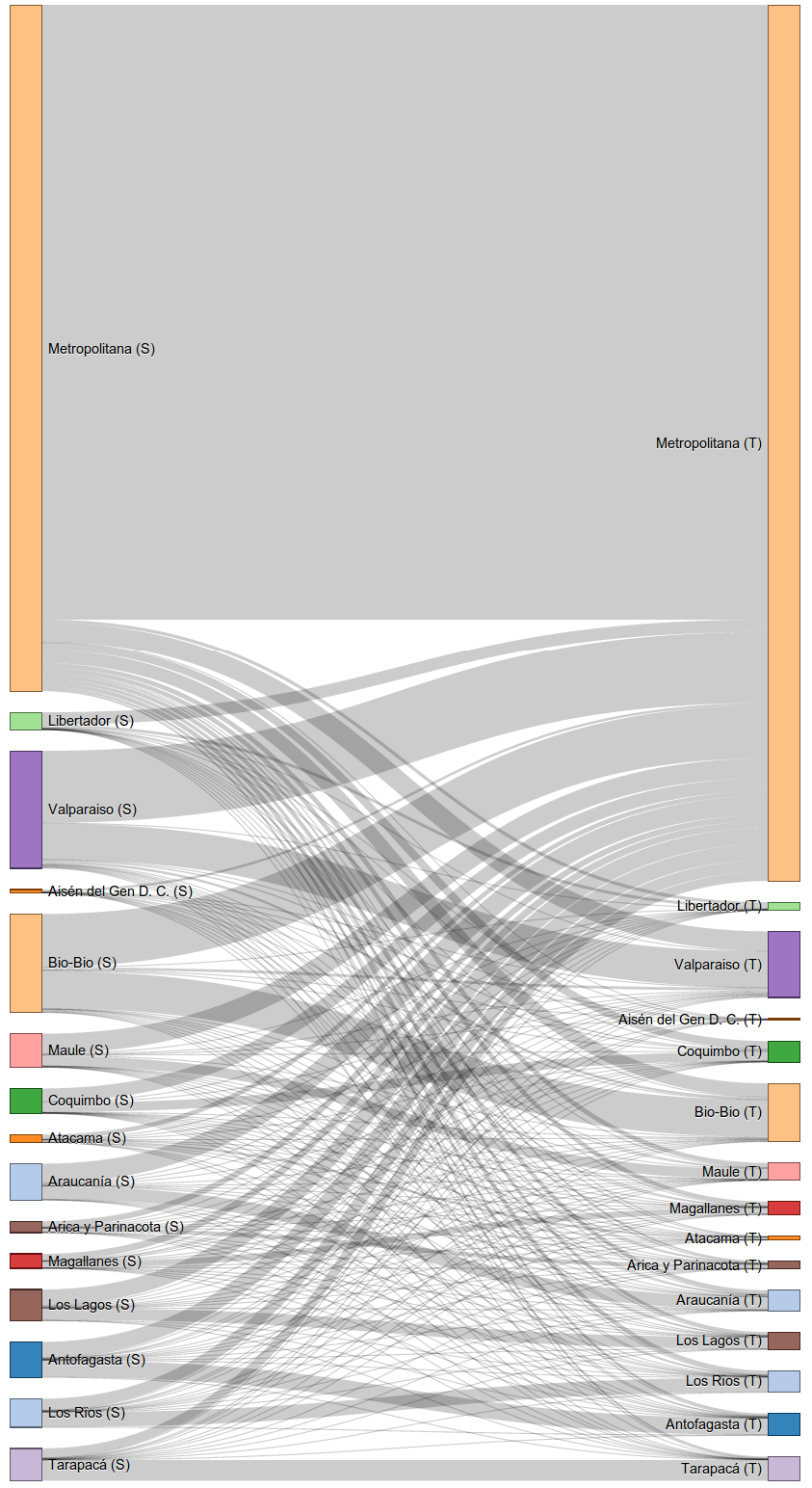}
\caption{Diagrama de flujo de interacciones entre regiones de Chile. Cada región es un rectángulo a cada lado del diagrama, cuyo tamaño es proporcional a la cantidad de información que emite (izquierda) y recibe (derecha). Fuente: elaboración propia.}
\label{fig:location_interactions}
\end{figure}

La Figura \ref{fig:location_interactions} muestra  un diagrama de flujo que representa la matriz origen-destino de menciones, respuestas y retweets entre regiones de Chile. Las regiones en la parte izquierda del gráfico envían tweets a las regiones de la parte derecha del gráfico. 
Ahora bien, el diagrama refleja los valores absolutos en el flujo de información, donde se observa que la Región Metropolitana genera mucho más flujo que el resto de las regiones, lo cual es de esperar dado el sesgo poblacional. 
También se observa que el flujo entrante en la Región Metropolitana es mayor al flujo saliente. 
La Tabla \ref{table:interaction_ratios} explora este fenómeno, al mostrar la proporción de flujo entrante y saliente de cada región, y el porcentaje de la información entrante que proviene de la misma región. Se observa que la Región Metropolitana es la única región que recibe más información de la que emite, aunque no es necesariamente la región en la que la mayoría de la información entrante viene de sí misma. Dicho puesto corresponde a la I Región, donde un $85.96\%$ de las interacciones recibidas provienen de sí misma. En promedio, el radio de entrada/salida de información es $0.67 \pm 0.21$ y el porcentaje de interacciones intra-región es $58.49\% \pm 14.06\%$. 

\begin{table}[htbp]
\centering
\footnotesize
\begin{tabular}{lcc}
\toprule
Reg. & Proporción In/Out & $\%$ de Contenido Intra-Región \\
\midrule
RM & $1.28$ & $70.11\%$ \\
XII & $0.90$ & $46.89\%$ \\
IV & $0.85$ & $43.72\%$ \\
XIV & $0.77$ & $67.26\%$ \\
I & $0.75$ & $85.96\%$ \\
XV & $0.68$ & $52.73\%$ \\
II & $0.62$ & $77.09\%$ \\
VIII & $0.59$ & $63.91\%$ \\
IX & $0.57$ & $62.75\%$ \\
V & $0.56$ & $54.70\%$ \\
X & $0.56$ & $66.78\%$ \\
III & $0.54$ & $57.39\%$ \\
VII & $0.53$ & $58.68\%$ \\
VI & $0.47$ & $35.61\%$ \\
XI & $0.36$ & $33.77\%$ \\
\bottomrule
\end{tabular}
\caption{Proporciones de entrada y salida de información para cada región del país, y porcentaje de flujo intra-regiones. Fuente: elaboración propia.}
\label{table:interaction_ratios}
\end{table}

\subsection{Tweets con Ubicación Geográfica}

Del total de tweets recolectados, $55898$ ($7.27\%$) contienen coordenadas geográficas. 
De dichos tweets, $53.26\%$ se encuentran en la área urbana de Santiago (que incluye todas las comunas de la provincia de Santiago y algunas comunas de las provincias de Cordillera y Maipo). A su vez, la mediana de latitud y longitud para estos tweets es $(-33.45, -70.67)$, exactamente las coordenadas de la capital, Santiago de Chile\footnote{Ver \url{https://es.wikipedia.org/wiki/Santiago_de_Chile}.}. 

\begin{figure}[htbp]
\centering
\includegraphics[width=0.75\textwidth]{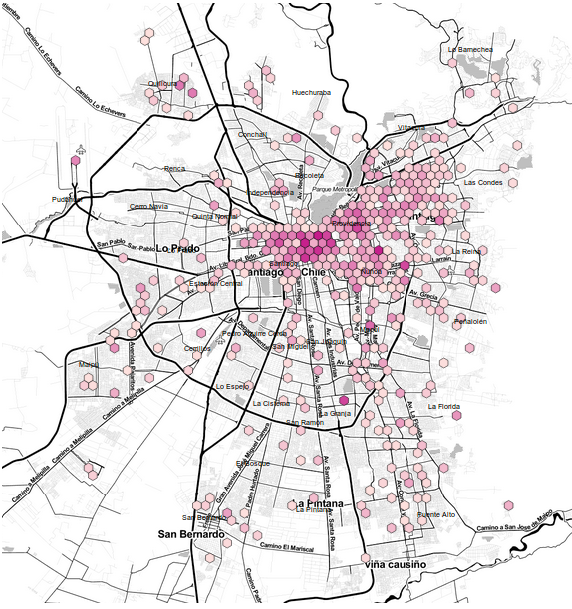}
\caption{Mapa de Santiago con distribución de casillas hexagonales. Cada casilla contiene al menos $20$ tweets, y mientras más oscuro sea su color, mayor es la cantidad de tweets que contiene. Fuente: elaboración propia. Las imágenes territoriales son de \emph{Stamen Design} y son utilizadas bajo licencia \emph{Creative Commons}.}
\label{fig:santiago_tweets}
\end{figure}

Estos tweets presentan un fuerte sesgo hacia la capital del país, por lo que esta sección se centrará en la ciudad de Santiago, al notar que los tweets asociados al resto del país son muy pocos. La Figura \ref{fig:santiago_tweets} muestra la distribución de tweets a través de casillas hexagonales con al menos $20$ tweets. El patrón que se observa parece corresponder con la \emph{densidad tecnológica} de la ciudad, puesto que comunas que tienen un alto grado de ingreso y con uso de suelo no solamente residencial tienen una mayor cantidad de tweets. Esto puede estar relacionado con la conclusión de la Sección \ref{sec:bios}, donde se menciona que el nivel educacional de la población virtual es alto.

\section{Discusión}

\subsection{Implicaciones}

No solamente el sesgo poblacional del país se ve reflejado (a través de la representatividad espacial), el centralismo que aqueja al país también está presente en Twitter: que la única región que reciba más información de la que emite sea la Región Metropolitana es un fuerte indicador de centralismo.

Además, existe sesgo hacia la población masculina. Si bien un porcentaje significativo de la población no ha sido clasificado, es razonable pensar que en ese grupo se mantienen las proporciones del grupo clasificado, basándonos en la literatura previa que ha encontrado un sesgo hacia los hombres en otros países \cite{mislove2011understanding}.

Por otro lado, es sabido que la población en redes sociales no es representativa en términos demográficos puesto que no toda la población tiene acceso a Internet. Dada la naturaleza del evento analizado es esperable un sesgo hacia la población mayor de edad. La alta participación de estudiantes y de profesionales indica que el nivel educacional de la población virtual participante en \texttt{\#municipales2012} es alto. Esto puede verse reflejado en el gráfico de Santiago con tweets geolocalizados, donde los tweets provienen en mayor proporción desde sectores con más recursos.

\subsection{Aplicaciones y Trabajo Futuro}

El resultado de este trabajo se ha utilizado como base para un sistema que predice las ubicaciones sobre las que habla un tweet \cite{graells2013santiago}, con el fin de crear una interfaz de usuario que provea diversidad geográfica en sus contenidos. 

Como trabajo futuro se propone estudiar el comportamiento de los registros de usuarios, con el fin de entender cuales \emph{peaks} son significativos y cuales son meramente producto del azar, y también analizar la distribución geográfica de tweets en otras ciudades de Chile, para estudiar con mayor profundidad la correlación entre densidad tecnológica y uso de Twitter con una muestra más grande.

\subsection{Limitaciones}

Existe un alto número de cuentas no clasificadas en términos de ubicación y de sexo de las personas que las registran. Para mejorar estos resultados es posible utilizar modelos de lenguaje, que han sido aplicados con éxito en el modelamiento de lugares geográficos \cite{o2013modeling} y de características léxicas en los tweets \cite{rao2010classifying}. Sin embargo, dicho enfoque está fuera del contexto de este artículo.

\section{Conclusiones}

En este artículo se han discutido y presentado diferentes señales a nivel geográfico y de comportamiento de las personas que participaron en Twitter en el marco de las elecciones municipales llevadas a cabo en Chile el año 2012. De acuerdo a la primera pregunta de investigación planteada, 
\emph{¿Cómo se puede caracterizar la población virtual chilena en Twitter?}, 
se ha mostrado que utilizando una muestra acotada pero a partir de un evento relevante a nivel nacional es posible obtener una caracterización de la población. 
Asimismo, con respecto a la segunda pregunta, 
\emph{¿Cómo determinar qué es lo que caracteriza el contenido de cada región del país? }, 
se ha mostrado que la utilización del modelo vectorial para representar documentos permite encontrar lo que caracteriza a cada región. 

No obstante el quiebre de barreras geográficas que supone una red social virtual, nuestra discusión muestra que es necesaria la creación de políticas que permitan mayor diversidad en el uso de las tecnologías, puesto que para Twitter, al menos para eventos de importancia nacional en los cuales se espera que haya alta participación, la población virtual chilena participante no es representativa demográficamente, a pesar de serlo espacialmente. 
Por otro lado, dado el hecho de que la interfaz y el diseño de Twitter no incentive el acceso a contenidos con diversidad geográfica, ni promueva la interacción entre personas de diferentes localidades, es necesario que se propongan paradigmas e interfaces de usuario que motiven a las personas a expresarse y a encontrar el contenido que les interesa, sin que su visión del contenido se vea sesgado producto de las distribuciones de población en el mundo físico.

\section*{Agradecimientos}

A Daniela Alarcón, Ricardo Baeza-Yates, Diego Sáez-Trumper, Daniele Quercia y Ruth García-Gavilanes por el apoyo, las ideas, discusiones y sugerencias que han permitido el avance de este proyecto. A Lorena Valderrama y al revisor anónimo por su valioso aporte para mejorar este trabajo.


{
\small
\bibliographystyle{ieeetr}
\bibliography{egraells_ornitologia_virtual}
}

\vspace{1cm}

\textbf{Eduardo Graells-Garrido} es estudiante de doctorado en el Grupo de Investigación de la Web de Universitat Pompeu Fabra y practicante en Yahoo Labs en Barcelona, bajo la supervisión de Dr. Ricardo Baeza-Yates. Obtuvo el grado de Magister en Ciencias, mención Computación en la Universidad de Chile. Ha trabajado en San Francisco, para la startup Identified, y en Santiago de Chile en el Departamento de Ingeniería Civil Transporte de la Universidad de Chile, en el Centro de Investigación de la Web y en diversos proyectos de Computación Gráfica para empresas particulares en Chile. 

\end{document}